\font\bbf=cmbx12

\centerline{\bbf Electromagnetic Zero Point Field as Active Energy Source}
\centerline{\bbf in the Intergalactic Medium}
\bigskip
\bigskip
\bigskip
\centerline{Alfonso Rueda  and Hiroki Sunahata}
\centerline{California State University, Long Beach, CA  90840}
\centerline{E-mail: arueda@csulb.edu}
\bigskip
\centerline{Bernhard Haisch}
\centerline{Solar \& Astrophysics Laboratory, Lockheed Martin}
\centerline{3251 Hanover St., Palo Alto, CA 94304}
\centerline{E-mail: haisch@starspot.com}
\bigskip
\centerline{\it Revised version of invited presentation at}
\centerline{35th AIAA/ASME/SAE/ASEE Joint Propulsion Conference and Exhibit}
\centerline{20--24 June 1999, Los Angeles, CA}
\centerline{AIAA paper 99-2145}
\footnote{\phantom{1}}{Copyright $\copyright$ 1999 by the American Institute of Aeronautics
and Astronautics, Inc. All rights reserved.}

\bigskip
\bigskip
\centerline{\bf ABSTRACT} 
\bigskip
For over twenty years the possibility that the electromagnetic zero point field
(ZPF) may actively accelerate electromagnetically interacting
particles in regions of extremely low particle density (as those extant in intergalactic
space (IGS) with $n \le 1$ particle m$^{-3}$) has been studied and analyzed. This
energizing phenomenon has been one of the few contenders for acceleration of cosmic rays
(CR), particularly at ultrahigh energies. The recent finding by the AGASA
collaboration ({\it Phys. Rev. Lett.}, {\bf 81}, 1163, 1998) that the CR energy spectrum
does not display any signs of the Greisen-Zatsepin-Kuzmin cut-off (that should be present if
these CR particles were indeed generated in localized ultrahigh energies CR
sources, as e.g., quasars and other highly active galactic nuclei), may indicate the need
for an acceleration mechanism that is distributed throughout IGS as
is the case with the ZPF. Other unexplained phenomena that receive an explanation from
this mechanism are the generation of X-ray and gamma-ray backgrounds and the existence of
Cosmic Voids. However recently, a statistical mechanics kind of challenge to the classical
(not the quantum) version of the zero-point acceleration mechanism has been posed (de
la Pe\~na and Cetto, {\it The Quantum Dice}, 1996). Here we briefly examine the consequences
of this challenge and a prospective resolution.

\bigskip
\centerline{\bf INTRODUCTION}
\bigskip
The idea that the
vacuum may play a fundamental role in the early development and future evolution of the
universe has been ``in the air'' for a long time [1]. Both the Steady State and the old
(pre-inflation model) Big Bang cosmological scenarios needed to
invoke this idea in one way or another [2]. The Steady State proposed the perennial creation
of particles (protons and electrons, presumably from the Dirac vacuum of
particle-antiparticle pairs) throughout the universe. The old Big Bang model postulated a
localized and instantaneous generation of all the matter and energy of the universe
together with the subsequent gradual generation of space-time which sprang from literally
nothing. And the new Big Bang with its more sophisticated Inflationary Cosmological Model
started from the postulation of an unstable vacuum (the false vacuum) that collapsed,
giving rise to the simultaneous creation of matter-energy and space-time followed by an
inflation which produces a rapid expansion of space-time [3, 4, 5]. This inflation [5] was
brief but extremely dynamic, and, as new space was actually being created, the process gave
the illusion of being superluminal. Once this stage terminated, the universe entered its
much slower current Hubble expansion [1, 5, 6], the era in which we now are and have been
for the last
$12-20
\times 10^9$ years or so.

\bigskip
If the vacuum played such an active role in the initial stages of the universe, a
valid question would be whether it is still playing some active role in the present
expansion of the universe. Recent observational findings by two different teams of
astronomers point in this direction [7, 8]. It has been discovered that instead of
a decreasing rate of expansion by cosmological gravitational attraction, as was assumed for
many years, the universe actually experiences an accelerated rate of expansion. This
accelerated rate of expansion may be physically explainable in terms of an ongoing
involvement of the vacuum in energizing the expansion.

\bigskip
\centerline{\bf ACCELERATED EXPANSION AND COSMIC VOIDS}
\bigskip
There is a paucity of known mechanisms by which the vacuum
might yield some of its energy to steadily contribute to an accelerated expansion of the
universe. Recently however within a different astrophysical context a mechanism
that could accomplish this was proposed [9, 10, 11].
Rueda, Haisch and Cole [11] investigated a mechanism that could account for the phenomenon
of Cosmic Voids [12]. It is now known that the structure of the universe
presents a peculiar distribution of matter in which clusters and superclusters of galaxies
are found in ``sheets'' that surround ``voids'', i.e., large regions of space of typical
diameters around 100 Megaparsecs, which are practically devoid of matter and where no
galaxies are found [1]. The whole structure may be represented by a soap-foam-like model
where the particles and magnetic fields are mainly found on the sheets along with galactic
clusters and superclusters, surrounding enormous spaces in between in which much lower
particle densities and much smaller concomitant magnetic field intensities prevail.

\bigskip
It has been proposed [10, 11] that the ZPF of traditional
quantum electrodynamics (QED), via a well-known mechanism (discussed below), is
responsible for the effect. 
This mechanism seems also to be involved in contributing to other 
astrophysical phenomena: the X-ray and gamma-ray backgrounds [13] and the
acceleration of cosmic rays (CR), particularly at very high energies, 
$E  \ge 10^{17}$ eV (see [9] for an extensive review).
It is important to emphasize that when the ZPF
is applied as an energizing entity to produce the expansion associated with the Cosmic
Voids, automatically an expansion of the universe itself must necessarily be produced.

\bigskip
The ZPF acceleration mechanism expands the Voids by creating a
pressure imbalance, transferring energy to, and thereby increasing the pressure in,
those regions where matter densities happen to be comparatively lower in the IGS
plasma [10, 11]. As a consequence the low-density regions tend to expand and to expel the
trapped magnetic fields. It is well known in astrophysics that space plasma regions of
higher densities are colder, while regions of lower density are almost exclusively occupied
by a highly energetic (hot) plasma of electrons and ionized nuclei, mainly protons [15].
This temperature-density anticorrelation is what occurs in the Voids and it is also what is
happening in many places throughout IGS. Moreover, such a
distribution would be a natural outcome of the ZPF acceleration mechanism when combined
with ordinary radiative collisional cooling [9].

\bigskip
\centerline{\bf THE ZERO-POINT FIELD ACCELERATION MECHANISM}
\bigskip
The origins of this mechanism go back to the early work of Einstein and his immediate
collaborators [16, 17]. Einstein realized that when a gas of electromagnetically
interacting particles
\footnote{$^a$}{Einstein restricted his considertion to polarizable particles. However it
can be shown that in the case of monopolar particles the mechanism is even more effective
[9, 18]} is submitted to the action of a random electromagnetic background (e.g., the case
of thermal radiation) two simultaneous phenomena take place. Due to the action of the random
electromagnetic medium, electromagnetically interacting particles become energized,
steadily increasing their translational kinetic energy. They perform a random walk in
velocity space that takes them, on average, systematically away from the origin.
Simultaneously however, as their velocities increase the particles find themselves
submerged in a random electromagnetic medium that is viewed by each of them as
Doppler-shifted and thereby has lost its isotropy. This causes the random electromagnetic
background to appear distorted and thereby to produce a drag force on the particle that
is of a frictional character because it is exactly proportional to the velocity $\vec{v}$,

$$
\vec{F}=-A \left[ \rho(\omega,T)-{1 \over 3}\omega {\partial \rho(\omega,T) \over
\partial \omega}\right] {\vec{v} \over c}, \eqno(1)
$$
where $A$ is a positive constant. In a Hohlraum, at
equilibrium temperature $T, \ \rho (\omega,T)$ represents the volumetric spectral energy
density of the radiation:
$$
\rho(\omega,T) d\omega ={\hbar \omega^3 \over 2 \pi^2 c^3} {\rm Coth} \left(
{\hbar \omega \over 2kT}\right) d\omega ={\hbar \omega^3 \over 2 \pi^2 c^3} \left[ {1 \over
{\rm exp} (\hbar \omega / kT)-1} + {1 \over 2}\right] d\omega \ . \eqno(2)
$$
The first term in the last parenthesis represents the thermal part: a Planck distribution at
temperature $T$. It disappears at zero temperature $(T \rightarrow 0)$, leaving the last
part,
$$
\rho(\omega,0) d\omega = \rho_0(\omega) d\omega= {\hbar \omega^3 \over 2 \pi^2 c^3}d\omega
\eqno(3)
$$
This is the spectral energy density of the ZPF. It originates in quantum theory from the
harmonic oscillator behavior of the individual cavity modes. At $T=0$, each individual
cavity mode behaves as a quantized harmonic oscillator with minimum, or zero-point, ground
state energy $\hbar \omega/2$. When the energy in each oscillator is multiplied by the
density of modes per unit volume $(\omega^2/\pi^2 c^3)$, one obtains the ZPF
spectral energy density above.

\bigskip
It can be shown that such ZPF is also present in free space, and cogent arguments can
be given for its reality [19, 20]. Observe however, that when the temperature is set to zero
(or close to zero) $\rho(\omega,T)d\omega \rightarrow \rho_0(\omega)d\omega$ and because of
the $\omega^3$ dependence, the Einstein-Hopf drag disappears. This last fact, first realized
by Boyer [21], is at the basis of the ZPF acceleration mechanism. So, under circumstances
in which there are negligible particle collisions and negligible ambient radiation fields
other than the the ZPF, when the temperature is low or negligible,
particles are still translationally energized by the random background ZPF radiation.
{\it But the Einstein-Hopf drag due to the ZPF is zero.} In the original
formulation of a ZPF acceleration mechanism [21, 16], it was assumed for simplicity that the
translational displacement of the particle was restricted to a single dimension (say, the
$x$-axis) and that the internal dipole vibrated along a single direction (say the
$z$-axis). These restrictions were removed by one of us when proposing this concept as a
mechanism for the actual acceleration of cosmic ray (CR) particles in IGS [14, 9].  Soon
after it was realized [18] that monopolar particles could also be accelerated by the ZPF,
but in a much more effective manner than polarizable particles. This conforms with the
well-known observational constraint on CR acceleration mechanisms that restricts the
acceleration to fully ionized nuclei [9]. Another well-known constraint is that electrons
appear in CR only at the very low energies, $E \le 10^{12}$ eV, and not beyond. This could
be explained [22] by the ultrarelativistic Zitterbewegung that, because of a time dilation
effect, decorrelates the actions of the electric and of the magnetic fields in the
Einstein-Hopf mechanism. This therefore prevents the acceleration of electrons to ultrahigh
CR energies. This decorrelation does not take place in the case of the much more massive
and sturdy protons that, if allowed, can be carried up to the highest CR energies beyond
$10^{20}$ eV [9].

\bigskip
The ZPF CR acceleration mechanism can  be derived in a quantum way [23, 24]. It was found
that it occurs in a time-symmetric or Wheeler-Feynman version of QED. But acceleration does
not occur in the more ordinary time-unidirectional version of QED [24, 9]. However, as the
time-symmetric QED version and the time-unidirectional version are equivalent (as long as
certain initial boundary conditions are assumed for the radiation in space-time and as
those conditions seem to hold in the original universe [25]) there is no clear reason for
taking one or the other versions of QED: no reason other than the fact that we are more
used to the time-unidirectional version.

\bigskip
It is moreover very interesting to mention that
both the Wheeler-Feynman version of QED and the classical stochastic theory give exactly
the same final form for the translational kinetic energy rate of growth $\Omega$, namely

$$
\Omega={3 \over 5\pi} (\Gamma \omega_0 )^2 \left( {\hbar \omega_0 \over m c^2} \right)
(\hbar \omega_0) \ \omega_0 
\eqno(4)
$$
where $\Gamma$ is the Abraham-Lorentz parameter ($\Gamma=2e^2/3mc^3$),
with $e$ the charge and $m$ the mass of the particle (proton). The frequency $\omega_0$ is a
parameter that depends on other considerations, e.g., what entity really performs the
oscillations, whether the whole proton or some component (like quarks, proton vibration
modes, etc.) inside the proton. In the most simplistic case $\omega_0$ comes to be half
the proton Compton frequency $(\omega_0=mc^2/2\hbar)$, but this applies only under
unrealistic strictly subrelativistic considerations. In practice $\omega_0$ is taken as a
free-parameter to be phenomenologically fitted by observation.

\bigskip
The ZPF acceleration mechanism could be found to satisfy all standard CR observational
constraints [26, 9], certainly at energies $E \ge 10^{17}$ eV. But lower
energies could not be immediately excluded, though the situation there was somewhat less
certain [9]. The least that we can say then is that the mechanism, up to now, seems to be
one of the strongest contenders for CR acceleration at ultrahigh $E \ge 10^{17}$ eV
energies.

\bigskip
\centerline{\bf CHALLENGE TO THE ZPF ACCELERATION CONCEPT}
\bigskip
	Recently in their 1996 textbook on the theory of Stochastic Electrodynamics (SED) and
related theories, de la Pe\~na and Cetto [20] have challenged at least some aspects
of the ZPF acceleration concept. In a lucid reanalysis [27] of the Boyer derivation [21] of
the translational kinetic energy growth, they argue that if an arguably more realistic
non-Markovian stochastic process is assumed for the phenomenon in its classical form, no
systematic translational kinetic energy growth takes place. This would then rather fit,
according to de la Pe\~na and Cetto, the time-unidirectional version of the QED acceleration
mechanism that indeed yields no systematic translational kinetic energy growth [24, 9].
We discuss the de la Pe\~na and Cetto argument in the Appendix.
\bigskip
Formally, of course, and once the non-Markovian behavior is assumed, the argument of
de la P\~ena and Cetto seems faultless. The situation however is far from clear.
Time-symmetric QED [25] still gives the acceleration and there is no certainty at all that
even when a classical viewpoint is implemented, the process has to be non-Markovian, i.e.,
having some memory. Moreover, recent work of Cole [28] strongly suggests the thermodynamics
soundness of the ZPF acceleration mechanism both in its physical and in its astrophysical
context. This step is important. Before this there remained a thermodynamic challenge to ZPF
acceleration [29]. Apparently, ZPF acceleration seemed to violate standard interpretations
of the first and of the second laws of thermodynamics. It has been shown that this is indeed
not the case [28].

\bigskip
\centerline{\bf DISCUSSION}
\bigskip
Given the overwhelming astrophysical explanatory possibilities of the ZPF acceleration
mechanism --- to mention a few, accelerated cosmic expansion, ultrahigh energies CR, part of
the X-ray and the gamma-ray backgrounds, Cosmic Voids, etc. --- it is of paramount
importance to clarify the situation and decide if the de la Pe\~na and Cetto challenge is
or is not a surmountable difficulty.

\bigskip
There are also other important possibilities for the ZPF acceleration mechanism. If
valid, the mechanism should eventually provide a means to transfer energy, back and forth,
but most importantly forth [9], from the vacuum electromagnetic ZPF into a suitable
experimental apparatus. A more far-fetched but not trifling possibility is that a better
understanding of the Einstein-Hopf process, that accompanies ZPF acceleration, would lead
to an understanding of the recently proposed ZPF contribution to inertia [30], also
presumably to some means for influencing inertia, and by the Einstein Principle of
Equivalence, also gravity. This has very interesting prospective engineering applications.

\bigskip
\centerline{\bf ACKNOWLEDGEMENTS}
\bigskip
BH and AR acknowledge partial support from NASA research contract NASW-5050. AR and BH
acknowledge interesting exchanges with Prof. Daniel C. Cole (Boston University).

\vfill\eject
\centerline{\bf APPENDIX: Possible suppression of the secular ZPF
acceleration}  

\bigskip
De la Pe\~{n}a and Cetto [20] have reached the conclusion that the zero-point field may
not produce secular acceleration. 
In this Appendix, the arguments leading to this conclusion are discussed.

\bigskip\noindent{\bf 1. Dipole Oscillator Model}

\bigskip
The Boyer SED approach to the problem [21] is based upon the model
originally developed by Einstein and Hopf [16, 17]. In this model, that for simplicity we
follow here, it is assumed that the internal particle motion is in the $x$-direction and
the oscillator dipole vibrates along the $z$-axis. (An extension to fully three-dimensional
motions was obtained in [14].)

\bigskip
During a time interval $\delta t$, this oscillator experiences two forces 
due to electromagnetic radiation, namely the impulse $\Delta$
transferred to the dipole via the interaction with the fluctuating
field, and the force of resistance $Rp$ due to the anisotropy of the
field as seen by the moving particle. Then, if at time $t$, the
momentum of the $translational$ motion of the oscillator is $p$, after 
a short time $\delta t$, its momentum becomes $p + \Delta - Rp\delta
t$. Since in equilibrium the mean-square momentum has to be constant
in time, we get the equilibrium condition,

$$
\left<p^2\right> = \left<(p + \Delta - Rp\delta t)^2\right>,\eqno(A1)
$$
where the impulse $\Delta = \int F dt$, and the drag coefficient $R$ are 
given respectively by 

$$
\left<\Delta^2\right> = \left<\left(\int^{t+\delta t}_{t} dt\, ez 
                        {\partial E_z \over \partial x}\right)^2\right> 
                      = {4 \tau \pi^4 c^4 \over 5 \omega^2} \rho^2 
                        (\omega, T) \delta t,\eqno(A2)
$$
and

$$
R = {6 \pi^2 c \tau \over 5m} \left( \rho - {1 \over 3} \omega {\partial
\rho \over \partial \omega} \right) . \eqno(A3)
$$

\bigskip
Equation (A1) , when expanded and neglecting the term of second order in
$\delta t$ yields

$$
\left< \Delta^2 \right> + 2 \left< p\Delta \right> - 2R \left< p^2 \right> \delta t
-2R \left< p\Delta \right> \delta t = 0. \eqno(A4)
$$
At $T=0$, however, there is no drag force since $\rho_0$ is
Lorentz-invariant, so that the above equation yields 

$$
\left< \Delta^2 \right>_0 + 2 \left< p\Delta \right>_0 = 0. \eqno(A5)
$$
This suggests that since $\langle \Delta^2 \rangle_0\, \neq 0$ due to
the presence of the zero-point field, the momentum $p$ and the
fluctuation  $\Delta$ have to be correlated, contrary to the
assumption $\langle p \Delta \rangle=0$ correctly made by Einstein and
his coworkers, but only for the case of pure thermal radiation. It is
here where de la Pe\~{n}a and Cetto disagree with Boyer [21] who assumed
even when the ZPF is present that $\langle p \Delta\rangle=0$
always. So, de la Pe\~{n}a and Cetto sensibly claim that the fluctuation at a
given time is not independent of past ones, i.e., the process $\delta
p = p - \bar{p}$ is not Markovian and the system acquires a certain
degree of memory in its interactions with the ZPF.  

\bigskip
Now let us combine (A4) with (A5) 
using the approximation $\langle p \Delta \rangle \approx \langle p
\Delta \rangle_0$ (which is justified because the thermal component of
the field is not expected to contribute significantly to the
correlation $\langle p \Delta \rangle_0$ ) to yield 

$$
\left< \Delta^2 \right> - \left< \Delta^2 \right>_0 
       = 2R \left< p^2 \right> \delta t - R \left< \Delta^2 \right>_0 \delta t. 
\eqno(A6) 
$$
It can be shown and is well-known to experts dealing with the model of 
Einstein and Hopf in SED that the stochastic average of the square of
the fluctuating impulse $\langle \Delta^2 \rangle_0$ is of order
$\delta t$. Thus, the last term is of order $(\delta t)^2$ and can be
neglected. Hence the equation simplifies to 

$$
\left< \Delta^2 \right> - \left< \Delta^2 \right>_0 
       = 2R \left< p^2 \right> \delta t. \eqno(A7)
$$
The first term includes both the thermal and zero-point fluctuations
and reduces to $\langle \Delta^2 \rangle_0$ at $T=0$. The drag force
on the right hand side is also zero at $T=0$ due to the $\omega^3$
dependence of the $\rho_0$. Since both sides reduce to zero at $T=0$,
it can be argued that $\langle \Delta^2 \rangle_0$ is no longer of the 
form $const \times \delta t$ that was responsible for the steady
translational kinetic energy growth. Hence no acceleration of a free
particle due to the zero-point field. The averaged square of the
fluctuating impulse should be a constant as in standard theory.

\bigskip\noindent{\bf 2. Quantum analysis}

\bigskip
Our quantum work [9, 23, 24] does not entirely support this De la Pe\~na and Cetto
conclusion. We find  that the secular acceleration disappears only under ordinary
time-unidirectional QED. However, the secular acceleration mechanism
is resurrected and in full force (with a much more detailed
algebraic expression that reduces in a suitable limit to the standard
classical case) under the Wheeler-Feynman time-symmetric form of
QED. A detailed discussion of this is found in the Appendix of
[9]. Our new exploration of this subject of the secular
acceleration mechanism involves among other things a reanalysis of
these results and a comparison of the classical Markovian case of the 
traditional ZPF SED secular acceleration case, the non-Markovian
counterpart of De la Pe\~na and Cetto as well as the corresponding time-asymmetric and
time-symmetric QED versions. We intend to pursue (and have proposed in
a NASA research proposal with D.C. Cole as PI) to perform an
experimental approach to the problem in order to check if the secular
acceleration can be validated for example in a Paul trap.

{

\bigskip
\parskip=0pt plus 2pt minus 1pt\leftskip=0.25in\parindent=-.25in

\centerline{\bf{REFERENCES}}

\medskip
[1] B\"orner, G. {\it The Early Universe} (Springer-Verlag, Heidelberg, 1988)
and references therein.

\medskip
[2] Weinberg, S. {\it Gravitation and Cosmology} (Wiley, New York,
1972) and references therein. 

\medskip
[3] Gliner, E.B., {\it Sov. Phys. JETP} {\bf 22}, 378 (1965).

\medskip
[4] Sato, K., {\it Mon. Not. R. Astron. Soc.} {\bf 195}, 487 (1981).

\medskip
[5] Guth, A., {\it Phys. Rev. D} {\bf 23}, 347 (1981).

\medskip
[6] Linde, A.D., {\it Phys. Lett.} {\bf 108B}, 389 (1982)

\medskip
[7] Perlmutter, S. et al, LBNL, preprint 41801
(1998) and for all relevant detailed updated information: www-supernova.lbl.gov

\medskip
[8] Perlmutter, S. et al., {\it Nature} {\bf 391}, 51 (1998) and references therein.

\medskip
[9] Rueda, A., {\it Space Science Reviews} {\bf 53},  223-345 (1990) and references therein.

\medskip
[10] Rueda, A., {\it Phys. Lett. A} {\bf 147}, 423 (1990).

\medskip
[11] Rueda, A., Haisch, B. \& Cole, D.C. {\it Astrophys. J.} {\bf 445},  7 (1995).

\medskip
[12] De Lapparent, V., Geller, M.J. \& Huchra, J.P., {\it Astrophys. J.} {\bf
302}, L1 (1986), and references therein.

\medskip
[13] Rueda, A., in {\it The Galactic and Extragalactic Background
Radiation}, IAU Symposium 139, Heidelberg, S. Bowyer and C. Leinert , eds.,  (Kluwer,
Dordrecht, 1990) pp 424-425.

\medskip
[14] Rueda, A., {\it Nuovo Cimento A} {\bf 48}, 155 (1978).

\medskip
[15] Ostriker, J., in Arons, J., McKee, C. \& Max, C., (eds.), {\it Particle Acceleration
Mechanisms in Astrophysics} (AIP, New York, 1979) p. 357.

\medskip
[16] Einstein, A. \& Hopf, L., {\it Ann. Phys. (Leipzig)} {\bf 33}, 1105 and 1096 (1910).

\medskip
[17] Einstein, A. \& Stern, O., {\it Ann. Phys. (Leipzig)} {\bf 40}, 551 (1913).

\medskip
[18] Rueda, A. {\it Phys. Rev A} {\bf 23}, 2020 (1981).

\medskip
[19] Brody, T., {\it The Philosphy Behind Physics} (Springer Verlag, Heidelberg 1993) 
L. de la Pe\~na \& P.E. Hodgson (eds.).

\medskip
[20] de la Pe\~na, L. \& Cetto, A.M. {\it The Quantum Dice} (Kluwer, Dordrecht, 1996).

\medskip
[21] Boyer, T. H., {\it Phys. Rev.} {\bf 182}, 1374 (1969).

\medskip
[22] Rueda, A. and Cavalleri, G., {\it Nuovo Cimento C} {\bf 6}, 239 (1983).

\medskip
[23] Rueda, A., {\it Phys. Rev. A} {\bf 30}, 2221 (1984).

\medskip
[24] Rueda, A. {\it Nuovo Cimento B} {\bf 96}, 64 (1986).

\medskip
[25] Davies, P.C.W. {\it The Physics of Time Asymmetry} (Univ. Calif. Press, Berkeley,
1974) and refernces therein.

\medskip
[26] Rueda, A. {\it Nuovo Cimento C} {\bf 6}, 523 (1983).

\medskip
[27] Ref. [20] pp. 147--152 and pp. 252--253.

\medskip
[28] Cole, D.C. {\it Phys. Rev. E} {\bf 51}, 1663 (1995)

\medskip
[29] Ref [9], p. 299 fff.

\medskip
[30] Rueda, A. \& Haisch, B.,{\it  Found. Phys.}, {\bf 28}, 1057 (1998); also
{\it Physics Letters A}, {\bf 240}, 115 (1998).

}
\bye